\documentclass[preprint,aps,showpacs,preprintnumbers,amsmath,amssymb,nofootinbib,epsf,epsfig]
{revtex4}

\usepackage{epsfig}
\begin{document}

\begin{flushright}
\end{flushright}


\newcommand{\be}{\begin{equation}}
\newcommand{\ee}{\end{equation}}
\newcommand{\bea}{\begin{eqnarray}}
\newcommand{\eea}{\end{eqnarray}}
\newcommand{\nn}{\nonumber}
\def\ds{\displaystyle}
\def\s1{\hat s}
\def\para{\parallel}
\newcommand{\mrm}[1]{\mathrm{#1}}
\newcommand{\mc}[1]{\mathcal{#1}}
\def\CP{{\it CP}~}
\def\cp{{\it CP}}
\def\ml{m_\mu}
\title{\large Effect of scalar leptoquarks on the rare decays of $B_s$ meson}
\author{  Rukmani Mohanta }
\affiliation{
School of Physics, University of Hyderabad, Hyderabad - 500 046, India }

\begin{abstract}
We study the effect of scalar leptoquarks on some rare decays of $B_s$ mesons involving the quark level
transition $b\rightarrow s$ $l^{+} l^{-}.$ In particular we consider the decays $B_{s} \rightarrow \mu^{+} \mu^{-}$,
$\bar B_d^0 \to X_s \mu^+ \mu^-$ and
$B_{s} \rightarrow \phi$ $\mu^{+} \mu^{-}$. The leptoquark parameter space is constrained using the recently measured branching
ratio of the $B_{s} \rightarrow \mu^{+} \mu^{-}$ process at LHCb and CMS experiments. Using such parameters we obtain the
branching ratio, forward backward  asymmetry and the CP asymmetry parameters in the angular distribution of
$B_{s} \rightarrow \phi$ $\mu^{+} \mu^{-}$ process.

\end{abstract}

\maketitle

\section{Introduction}

The standard model (SM) of electroweak interaction is very successful
in explaining the observed data so far and is further supported by the recent discovery of a Higgs-like
boson in the mass range of 126 GeV. But still there are many reasons to believe that
it is not the ultimate theory of nature, rather some low energy limit of
some more fundamental theory whose true nature is
not yet well understood. It is therefore an ideal time to test the predictions of the standard model
more carefully and try to identify the nature of physics beyond it.
If there would be  new physics (NP) at the TeV scale associated with the hierarchy problem, it is natural
to expect that it would first show up in the flavor sector and in this context the rare decays
of $B$ mesons induced by flavor changing neutral current (FCNC) transitions play a very crucial role.
The FCNC transitions are one-loop suppressed in the SM and thus provide an excellent
testing ground to look for possible existence of new physics.

In this paper we would like to investigate some  rare decay modes of $B_s$ meson using the scalar leptoquark (LQ) model. The study of $B_s$ meson has attracted a lot of attention in recent times as large number of
$B_s$ mesons are produced in the LHCb experiment and this would open up
the possibility to study the rare decays of $B_s$ meson with high statistical precision. The most important and sought  after rare decay mode  is the $B_s \to \mu^+ \mu^-$ process mediated by the FCNC
transition $b \to s$, has been recently observed by the
LHCb \cite{lhcb1} and CMS \cite{cms} collaborations. This mode is very interesting as it is theoretically very clean and
highly suppressed in the standard model and hence well suited for constraining the new physics
parameter space. Another important rare decay channel mediated by the quark level transition $ b \to s \mu^+ \mu^-$ is the inclusive decay process $\bar B_d^0 \to X_s \mu^+ \mu^-$. The integrated branching ratio for this process has been measured by both Belle \cite{belle-1} and BaBar \cite{babar-1} collaborations. It is expected that in  the low $q^2$ region (1 GeV$^2 \leq q^2 \leq 6$ GeV$^2$ ) as well as in the high $q^2$ region ($q^2 \geq $ 14.4 GeV$^2$) the theoretical predictions are dominated by perturbative contributions and hence a theoretical precision of order 10$\%$ is
in principle possible \cite{hurth}. We will use the measured branching ratios of these processes to constrain the leptoquark
parameters and subsequently apply these parameters to study the semileptonic rare decay mode $B_s \to \phi \mu^+ \mu^-$.

Leptoquarks are color-triplet bosons that can couple to a quark and a lepton at the same time and can occur
in various extensions of the SM \cite{lepto}.  Scalar leptoquarks are expected to exist at the TeV scale in extended  technicolor models \cite{lepto1} as well as  in models of quark and lepton compositeness \cite{lepto2}.
The general classification of leptoquark models are discussed in \cite{lepto3} and the phenomenology of
scalar leptoquarks have been studied extensively in the literature \cite{lepto4,wise,lepto5}.
Here, we will consider the model where leptoquarks can couple only to a pair of quarks and leptons and thus
may be inert with respect to proton decay. In such cases, proton decay bounds would not apply and
leptoquarks may produce signatures in other low-energy phenomena \cite{wise}.

The paper is organized as follows. In section II we briefly discuss the effective Hamiltonian describing the process $b \to s l^+ l^-$. The new contributions arising due to the exchange of scalar leptoquark are presented in section III.
We present the rare decay modes $B_s \to \mu^+ \mu^-$  and $\bar B_d^0 \to X_s \mu^+ \mu^-$ in sections IV and V respectively and obtain the constraints on leptoquark parameters. The decay mode $B_s \to \phi \mu^+ \mu^- $ is discussed in Section VI and section VII contains the Conclusion.

\section{Effective Hamiltonian for $b \to s l^+ l^- $ process in the Standard Model}

Within the standard model the effective Hamiltonian describing the quark level transition is given as \cite{buras1}
\bea
{\cal H}_{eff} &=& - \frac{ 4 G_F}{\sqrt 2} V_{tb} V_{ts}^* \Bigg[\sum_{i=1}^6 C_i(\mu) O_i +C_7 \frac{e}{16 \pi^2} \Big(\bar s \sigma_{\mu \nu}
(m_s P_L + m_b P_R ) b\Big) F^{\mu \nu} \nn\\
&&+C_9^{eff} \frac{\alpha}{4 \pi} (\bar s \gamma^\mu P_L b) \bar l \gamma_\mu l + C_{10} \frac{\alpha}{4 \pi} (\bar s \gamma^\mu P_L b)
\bar l \gamma_\mu \gamma_5 l\Bigg]\;,\label{ham}
\eea
where   $G_F$ is the Fermi constant and $V_{q q'}$ are the Cabibbo-Kobayashi-Maskawa (CKM)
matrix elements, $\alpha$ is the fine structure constant, $P_{L,R} = (1 \mp \gamma_5)/2$ and $C_i$'s are the Wilson coefficients. The values of the Wilson coefficients are calculated at the next-to-next-leading order (NNLO) by matching the full theory to the effective theory
at the electroweak scale and subsequently solving the renormalization group equation to run them down to the b-quark mass scale i.e., $\mu_b=4.8$ GeV \cite{buras2}.

The coefficient $C_9^{eff}$ contains a perturbative part  and a
resonance part which comes
from the long distance effects due to the conversion of the real
$c \bar c$ into the lepton pair $l^+ l^-$. Thus, $C_9^{eff}$ can be
written as
\be
C_9^{eff}=C_9+Y(s)+C_9^{res}\;,
\ee
where $s=q^2$ and the function $Y(s)$ denotes the perturbative part coming
from one loop matrix elements  of the four quark operators and
is given in Ref. \cite{buras1}.
The long distance resonance effect is given as \cite{res}
\bea
C_9^{res}= \frac{3 \pi}{\alpha^2}(3 C_1+C_2+3C_3+C_4+3C_5+C_6)\sum_{V_i=J/\psi,
\psi^\prime} \kappa\frac{m_{V_i} \Gamma(V_i \to l^+ l^-)}{m_{V_i}^2 -s
-i m_{V_i}\Gamma_{V_i}}\;,
\eea
where the phenomenological parameter $\kappa$ is taken as 1.7 and 2.4 for the two lowest two $\bar c c$ resonances $J/\psi$ and
$\psi'$ \cite{buras2}.

\section{New Physics Contributions due to Scalar Leptoquark exchange}

In the leptoquark model the effective Hamiltonian describing the process $b \to s l^+ l^-$ will be modified due
to the additional contributions arising from the exchange of leptoquarks. Here, we will consider the minimal renormalizable scalar leptoquark models \cite{wise}, where the standard model is augmented only by one additional
scalar representation of $SU(3) \times SU(2) \times U(1)$ and which  do  not allow proton decay at the tree
level. It has been shown in \cite{wise} that there are only two models which can
satisfy this requirement.  In these models the leptoquarks
have the representation as  $X=(3,2,7/6)$ and $X=(3,2,1/6)$ under the  $SU(3) \times SU(2) \times U(1)$ gauge group.
Our aim  here is to consider these scalar leptoquarks which potentially contribute to the $b \to s \mu^+ \mu^-$ transitions and constrain the underlying couplings from experimental data on $B_s \to \mu^+ \mu^-$ and $\bar B_d^0 \to X_s \mu^+ \mu^-$.
Although the decay modes $\bar B_d^0 \to \bar K^0 \mu^+ \mu^-$ and $\bar B_d^0 \to K^{* 0} \mu^+ \mu^-$ are also mediated by the same quark level transition $b \to s \mu^+ \mu^-$, we do not consider the measured branching ratios of such processes to constrain the NP parameter space as these measurements involve additional uncertainties due to the form factors.
However, we will comment on the recent observation of several anomalies on  angular observables in the rare decay $B \to
K^{*0} \mu^+ \mu^-$ by the LHCb collaboration \cite{lhcb13}.

Now we consider all possible renormalizable  interactions of such leptoquarks with SM matter fields consistent with the SM gauge symmetry in the following subsections.
\subsection{Model I: $X= (3,2,7/6)$ }
In this model the interaction Lagrangian for the coupling of scalar leptoquark $X= (3,2,7/6)$ to the fermion bilinears is given as \cite{wise}
\bea
{\cal L} = - \lambda_{u}^{ij}~ \bar u_R^i X^T \epsilon L_L^j - \lambda_e^{ij}~\bar e_R^i X^\dagger Q_L^j + h.c.\;,
\eea
where $i,j$ are the generation indices,  $Q_L$ and $L_L$ are the left handed quark and lepton doublets,
$u_R$ and $e_R$ are the right handed up-type quark and charged lepton singlets and $\epsilon=i \sigma_2$ is a $2\times 2$ matrix. More explicitly these multiplets can be represented as
\bea
X=
\left( \begin{array}{c}
 V_\alpha   \\
Y_\alpha
\end{array}
\right ), \; \;\:\:\:
L_L =
\left ( \begin{array}{c}
\nu_L\\
e_L \\
\end{array}
\right ) ,~~~~{\rm and}~~~~~
\epsilon=
\left( \begin{array}{cc}
 0 & 1  \\
-1~ & 0 \\
\end{array}
\right ).
\eea
After expanding the $SU(2)$ indices the interaction Lagrangian becomes
\bea
{\cal L}= -\lambda_u^{ij}~ \bar u_{\alpha R}^i ( V_\alpha e_L^j - Y_\alpha \nu_L^j )
-\lambda_e^{ij}~ \bar e_R^i \left (V_L^\dagger u_{\alpha L}^j + Y_\alpha^\dagger d_{\alpha L}^j \right )+h.c.\;.\label{lepto}
\eea
Thus, from Eq. (\ref{lepto}), one can obtain the contribution to the interaction Hamiltonian for
the  $b \to s \mu^+ \mu^- $ process
 after Fierz rearrangement as
\bea
{\cal H}_{LQ}&=& \frac{\lambda_\mu^{23} \lambda_\mu^{22 *} }{8 M_Y^2} [ \bar s \gamma^\mu (1-\gamma_5)b]
[\bar \mu \gamma_\mu(1+\gamma_5) \mu ]\equiv \frac{\lambda_\mu^{23} \lambda_\mu^{22 *}}{4 M_Y^2} \Big (O_9 +O_{10} \Big),
\eea
which can be written analogous to the SM effective Hamiltonian (1) as
\bea
{\cal H}_{LQ}=- \frac{  G_F \alpha}{\sqrt 2 \pi} V_{tb} V_{ts}^*(C_9^{NP} O_9 +C_{10}^{NP} O_{10})
\eea
with the new Wilson coefficients
\bea
C_9^{NP} = C_{10}^{NP} = - \frac{ \pi}{2 \sqrt 2 G_F \alpha V_{tb} V_{ts}^* }\frac{\lambda_\mu^{23} \lambda_\mu^{22 *}}{
M_Y^2}\;.\label{c10np}
\eea

\subsection{Model II: X= (3,2,1/6)}
Analogous to the previous subsection the interaction Lagrangian for the coupling of
$X=(3,2,1/6)$ leptoquark to the fermion bilinear can be given as
\bea
{\cal L}= - \lambda_d^{ij}~\bar{d}_R^i X^T \epsilon L_L^j + h.c.\;,
\eea
where the notations used are same as the previous case. Expanding the $SU(2)$ indices one can obtain
the interaction Lagrangian as
 \bea
{\cal L} = - \lambda_d^{ij}~ \bar d_{\alpha R} (V_\alpha e_L^j-Y_\alpha \nu_L^j) +h.c.\;.
\eea
After performing the Fierz transformation the interaction Hamiltonian describing the process
$b \to s \mu^+ \mu^-$ is given as
\bea
{\cal H}_{LQ}&=& \frac{\lambda_s^{22} \lambda_b^{32*}}{4 M_V^2}[\bar s \gamma^\mu P_R b] [ \bar \mu \gamma_\mu (1- \gamma_5) \mu ]= \frac{\lambda_s^{22} \lambda_b^{32*}}{4 M_V^2}\left ( O_9^{'NP }-O_{10}^{'NP } \right )\;,
\eea
where $O_9'$ and $O_{10}'$ are  the four-fermion current-current operators  obtained from $O_{9,10}$ by making the replacement $P_L \leftrightarrow P_R$.
Thus, the exchange of the leptoquark $X=(3,2,1/6)$ gives new operators with the corresponding Wilson coefficients as
\bea
C_9^{'NP } = - C_{10}^{'NP } = \frac{ \pi}{2 \sqrt 2 ~G_F \alpha V_{tb}V_{ts}^*} \frac{\lambda_s^{22} \lambda_b^{32*}}{M_V^2}\;.\label{c10np1}
\eea
After obtaining the new physics contributions to the process $ b \to s \mu^+ \mu^-$, we will proceed the constrain the new physics parameter space using the recent measurement of $B_s \to \mu^+ \mu^-$.

\section{$B_s \to \mu^+ \mu^-$ decay process}

The rare decay process $B_s \to \mu^+ \mu^-$, mediated by the FCNC transition
$b \to s$ is strongly helicity suppressed  in the standard model. Furthermore, it
is very clean and the only nonperturbative quantity involved is
the decay constant of $B_s$ meson which can be reliably calculated by the well known
non-perturbative methods such as QCD sum rules, lattice gauge theory etc.
Therefore, it is believed to be one of the most powerful tools to look for new physics beyond the standard model.
This process has been very well studied in the literature and in recent times also it has attracted a lot of attention
\cite{fleischer1,fleischer2,buras5,mu6,staub,kosnik}. Therefore, here we will quote  the important results.

The most general effective Hamiltonian
describing this process
\bea
{\cal{H}}_{eff}
= \frac{G_F \alpha }{\sqrt{2} \pi} V_{tb} V_{ts}^*
\Bigg[C_{10}^{eff}O_{10} + C_{10}' O_{10}' \Bigg],\label{hammu}
\eea
where $C_{10}^{eff} = C_{10}^{SM} + C_{10}^{NP}$ and $C_{10}'= C_{10}^{'NP}$.
The branching ratio for this process is given as
\be
{\rm BR}(B_s \to \mu^+ \mu^-) = \frac{G_F^2}{16 \pi^3} \tau_{B_s} \alpha^2 f_{B_s}^2 m_{B_s} m_{\mu}^2 |V_{tb} V_{ts}^*|^2
\left |C_{10}^{eff}-C_{10}'\right |^2 \sqrt{1- \frac{4 \ml^2}{m_{B_s}^2}}.
\ee
Now using $\alpha = 1/128$, $|V_{tb}V_{ts}^*|=0.0405 \pm 0.0008$, $f_{B_s}=227 \pm 8 $ MeV \cite{buras5}, $C_{10}^{SM}=-4.134$
\cite{fleischer2},
the particle masses and lifetime of $B_s$ meson from \cite{pdg} we obtain the SM branching ratio for this process as
\be
{\rm BR }(B_s \to \mu^+ \mu^-) = (3.29 \pm 0.19)\times 10^{-9},\label{brmu}
\ee
which is  consistent with the latest SM prediction ${\rm Br}(B_s \to \mu^+ \mu^-)=(3.23 \pm 0.23) \times 10^{-9}$ \cite{buras5}.
The branching ratio for this mode has recently been measured by both LHCb \cite{lhcb1} and CMS \cite{cms} collaborations.
Analyzing the data corresponding to an integrated luminosity of $1~{\rm fb}^{-1}$ at $\sqrt{7}$ TeV and $2~{\rm fb}^{-1}$ at $\sqrt{8}$ TeV
the LHCb collaboration obtained the time integrated branching ratio as
\be
{\rm BR}(B_s \to \mu^+ \mu^-)=(2.9_{-1.0}^{+1.1}) \times 10^{-9}\;.
\ee
The CMS collaboration \cite{cms}  also obtained  analogous result
\be
{\rm BR}(B_s \to \mu^+ \mu^-)=(3.0_{-0.9}^{+1.0}) \times 10^{-9}\;,
\ee
where they have used the data samples corresponding to integrated luminosities of $5$ and $20~{\rm fb}^{-1}$ at ${\sqrt s}=7$ and 8 TeV.
The weighted average of these two measurements yields
 \be
{\rm BR}(B_s \to \mu^+ \mu^-)=(2.95 \pm 0.71 ) \times 10^{-9}\;,
\ee
which is consistent with the latest SM prediction (\ref{brmu}), but certainly it
does not rule out the possibility of new physics in this mode.
While new physics can still affect this decay mode, but certainly
its contribution is not the dominant one.

However, as discussed in Ref . \cite{fleischer1}, in the experiment the time integrated untagged decay rate is measured, whereas in the above theoretical calculation the effect of meson oscillation is not taken
into account. Therefore, while comparing the SM prediction for $B_s \to \mu^+ \mu^-$ decay rate with
the experimental result one should take into account the  sizable width difference
$\Delta \Gamma_s$ between $B_s $ mass eigenstates. i.e.,
\be
y_s\equiv \frac{\Gamma_L^{(s)} - \Gamma_H^{(s)}}{\Gamma_L^{(s)} + \Gamma_H^{(s)}}= \frac{\Delta \Gamma_s}{2 \Gamma_s}
=0.087 \pm 0.014\;,
\ee
where $\Gamma_s = \tau_{B_s}^{-1}$ denotes the average $B_s$ decay width. Hence, the experimental result
is related to the theoretical prediction as
\be
{\rm BR^{th}}(B_s \to \mu^+ \mu^-)= \left [\frac{1-y_s^2}{1+{\cal A}_{\Delta \Gamma} y_s }\right ] {\rm BR}(B_s \to \mu^+ \mu^-)^{\rm exp}\;,
\ee
where the observable ${\cal A}_{\Delta \Gamma}$  equals $+1$ in the SM. Thus, using the experimental value of $y_s$ we
obtain the branching ratio in the standard model
\be
{\rm BR}(B_s \to \mu^+ \mu^-)^{\rm th}|_{SM} =(3.60 \pm 0.21)\times 10^{-9}\;.
\ee

We will now consider the effect of scalar leptoquarks in this mode. One can write the transition amplitude for this
process from Eq. (\ref{hammu}) as
\bea
A(B_s^0 \to \mu^+ \mu^-)=\langle \mu^+ \mu^- |{\cal H}_{eff}| B_s^0 \rangle
= - \frac{G_F}{ {\sqrt 2}~ \pi}V_{tb}V_{ts}^* \alpha f_{B_s} m_{B_s} m_\mu C_{10}^{SM} P ,
\eea
where
\be
P \equiv \frac{C_{10}-C_{10}'}{C_{10}^{SM}}=1+ \frac{C_{10}^{NP}-C_{10}^{'NP}}{C_{10}^{SM}}=1+r e^{i \phi^{NP}}\;,
\ee
with
\be
r e^{i \phi^{NP}}= (C_{10}^{NP}-C_{10}^{'NP})/C_{10}^{SM}\;,\label{bound}
\ee
denotes  the new physics contribution
and $\phi^{NP}$ is the relative phase between SM and the NP couplings. In general
 $P \equiv|P|e^{\phi_P}$ carries the CP violating phase $\phi_P$. The phases $\phi_P$ and $\phi^{NP}$ are related to
 each other by the relation
 \be
 \tan \phi_P= \frac{r \sin \phi^{NP}}{1+r \phi^{NP}}\;.
 \ee
As discussed in section III,  the exchange of the leptoquark $X(3,2,7/6)$  gives new contribution to $C_{10}$ and $X(3,2,1/6)$  gives additional contribution $C_{10}'$ the branching ratio
in both the cases will be
\be
{\rm BR}(B_s \to \mu^+ \mu^-)^{\rm th} = \left [\frac{1+ {\cal A}_{\Delta \Gamma}}{1-y_s^2}\right ]{\rm BR}^{SM}(1+r^2 -2 r \cos \phi^{NP})\;.
\ee
In the leptoquark model the observable ${\cal A}_{\Delta \Gamma} $ becomes \cite{fleischer1}
\be
 {\cal A}_{\Delta \Gamma}=\cos 2 \phi_P\;.
 \ee
\begin{figure}[htb]
\centerline{\epsfysize 2.5 truein \epsfbox{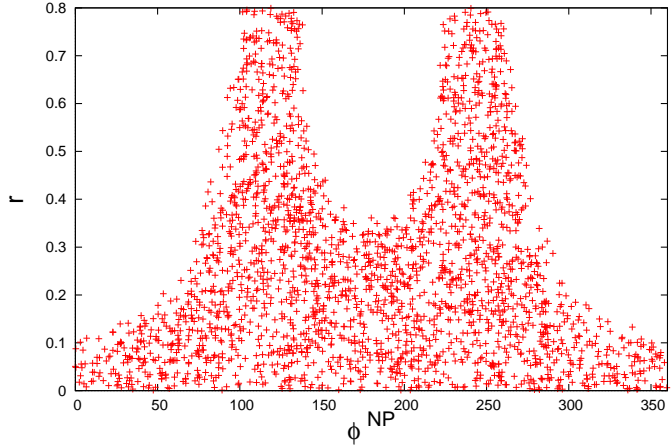}}
\caption{The allowed region in the $r-\phi^{NP}$ parameters space obtained from the
${\rm BR}(B_s \to \mu^+ \mu^-)$.}
\end{figure}
 In order to find the constrain on the combination of LQ couplings we
require that each individual leptoquark contribution  to the branching ratio does not exceed
the experimental result.
Now using the SM value from (\ref{brmu}), we show in Fig. 1 the allowed region
in $r-\phi^{NP}$ plane which is compatible with the $2\sigma$ range of
the experimental data. From the figure one can see that for $0\leq r \leq 0.1 $ the entire range for $\phi^{NP}$ is allowed, i.e.,
\bea
 0\leq r \leq 0.1\;, ~~~~{\rm for}~~~~0 \leq \phi^{NP} \leq 2 \pi \;.\label{r-bnd}
 \eea

\section{Analysis of $\bar{B}_d^0 \to X_s \mu^+ \mu^-$ mode}

Now we would like to constrain the NP couplings from the measured branching ratio of the inclusive decay
$\bar B_d^0 \to X_s \mu^+ \mu^-$. The integrated branching ratio for this process has been measured by both Belle \cite{belle-1}
 and BaBar \cite{babar-1} collaborations and the average value of these measurements in the two regions are
 \cite{hurth}
\bea
{\rm BR}(B_d^0 \to X_s \mu^+ \mu^-) &=& (1.60 \pm 0.50) \times 10^{-6}~~~~~{\rm low}~q^2\nn\\
&=&(0.44 \pm 0.12) \times 10^{-6}~~~~~{\rm high}~ q^2\;,\label{br-2}
\eea
where the low-$q^2$ and high-$q^2$  regions correspond to 1 GeV$^2 \leq q^2 \leq $ 6 GeV$^2$ and
$q^2\geq 14.4$ GeV$^2$, respectively.
The decay mode has been very well studied in the literature and here we are presenting only the
main results.
The differential branching ratio for  this  process in the standard model is given as \cite{alok-1}
\bea
\frac{d{\rm BR}}{ds_1}\biggr|_{\rm SM} &= & B_0 \frac{8}{3} (1-s_1)^2\sqrt{1- \frac{4t^2}{s_1}} \times \biggr[(2s_1+1)
\left(\frac{2t^2}{s_1} +1 \right ) |C_9^{eff}|^2\nn\\
&+& \left (\frac{2(1-4 s_1)t^2}{s_1} +(2 s_1+1)\right )  |C_{10}  |^2
+4 \left (\frac{2}{s_1}+1 \right )\left (\frac{2 t^2}{s_1}+1 \right )\left |C_7 \right |^2\nn\\
&+& 12 \left (\frac{2 t^2}{s_1}+ 1\right ){\rm Re}(C_7 C_9^{eff *} ) \biggr]\;,
\eea
where $t= m_\mu/m_b^{pole}$ and $s_1=q^2/(m_b^{pole})^2$. The normalization constant $B_0$ is related to ${\rm BR}(\bar B
\to X_c e \bar \nu_e)$ through
\bea
B_0 = \frac{3 \alpha^2 {\rm BR} (\bar B \to X_c e \bar \nu_e)}{32 \pi^2 f(\hat m_c)\kappa(\hat m_c)}
\frac{|V_{tb} V_{ts}^{*}|^2}{|V_{cb}|^2}\;,
\eea
where $\hat m_c = m_c^{pole}/m_b^{pole}$. $f(\hat m_c)$ is the lowest order phase space factor for the
$\bar B \to X_c e \bar \nu$ process, i.e.,
\bea
f(\hat m_c)= 1-8 \hat m_c^2 + 8 \hat m_c^6 - \hat m_c^8 -24 \hat m_c^4 \ln \hat m_c\;,
\eea
and the function $\kappa(\hat m_c)$ is the power correction to ${\rm BR}(\bar B \to X_c e \bar \nu)$, which
includes both the $O(\alpha_s)$ QCD corrections and the leading order  $(1/m_b^2)$ power corrections
\be
\kappa(\hat m_c)= 1- \frac{2 \alpha_s(m_b)}{3 \pi} g(\hat m_c) + \frac{h(\hat m_c)}{2 m_b^2}\;.
\ee
Here the two functions are given as
\bea
g(\hat m_c) &=& \left (\pi^2 - \frac{31}{4}\right )(1-\hat m_c)^2 + \frac{3}{2}\;,\nn\\
h(\hat m_c) &=& \lambda_1 + \frac{\lambda_2}{f(\hat m_c)}\left [ -9 + 24 \hat m_c^2 -72 \hat m_c^4
+72 \hat m_c^6 -15 \hat m_c^8 -72 \hat m_c^8 - 72 \hat m_c^4 \ln \hat m_c \right ],
\eea
where $\lambda_1$ and $\lambda_2$  are the kinetic energy and magnetic moment operators.

In the leptoquark model there will be additional contribution arising due to the exchange of leptoquarks
which will introduce the new couplings $C_9^{NP}$, $C_{10}^{NP}$, $C_9^{'NP}$ and $C_{10}^{'NP}$ as
discussed in section III. Including these NP contributions and neglecting the sub-leading terms which are
suppressed by $m_\mu/m_b$ and $m_s/m_b$, the branching ratio can be  given as

\bea
\left (\frac{d {\rm BR}}{d s_1 }\right )_{\rm Total}&=&\left (\frac{d {\rm BR}}{d s_1 }\right )_{\rm SM}
+B_0 \Big[\frac{16}{3} (1-s_1)^2 (1+2 s_1)[{\rm Re}(C_9^{eff} C_9^{NP *}+{\rm Re}(C_{10} C_{10}^{NP *}]\nn\\
&+& \frac{8}{3}(1-s_1)^2(1+2 s_1)\left [|C_9^{NP}|^2 +|C_{10}^{NP}|^2 +|C_9^{'NP}|^2+|C_{10}^{'NP}|^2\right ]
\nn\\
&+& 32 (1-s_1)^2 ~{\rm Re}(C_7 C_{10}^{NP *})  \Big]\;.\label{br-1}
\eea
For numerical evaluation we use the input parameters as $\hat m_c = 0.29 \pm 0.02$ \cite{ali}, BR$(\bar B \to X_c e \bar \nu)=(10.1 \pm 0.4 ) \%$ \cite{pdg},
$|V_{tb}V_{ts}^*|/|V_{cb}|=0.967 \pm 0.009$ \cite{ckmfitter} and the parameters $\lambda_1$ and $\lambda_2$ as
$\lambda_1=-(0.1 \pm 0.05)$ GeV$^2$  and $\lambda_2=0.12 $ GeV$^2$ \cite{hiller}.
With these parameters the branching ratio in the SM is found to be
\bea
{\rm BR}(\bar B \to X_s \mu^+ \mu^-) &=& (1.92) \pm 0.08)\times 10^{-6}~~~~{\rm low}~q^2\nn\\
&=& (0.38 \pm 0.01) \times 10^{-6}~~~~~{\rm high}~q^2\;.\label{br-3}
\eea
These predicted branching ratios are in agreement with the corresponding experimental values within
their 1-$\sigma$ range. To constrain the new physics couplings coming from the exchange of scalar  leptoquarks
$X(3,2,7/6)$ and $X(3,2,1/6)$, we assume only one type of leptoquark will contribute at a time. As discussed
in section III, in the presence of the leptoquark $X(3,2,7/6)$ only the NP couplings $C_9^{NP}$ and $C_{10}^{NP}$
will arise whereas for $X(3,2,1/6)$ the couplings $C_9^{'NP}$ and $C_{'10}^{NP}$ will contribute.
Furthermore, as shown in Eqs. (9) and (13) the magnitudes of these couplings in each case will be same.
With the additional assumption that
these two couplings will have the same phase $\phi^{NP}$ and neglecting the small phase difference
 between $C_9^{eff}$ and $C_{10}^{NP}$ we obtain the constraint equations for these NP couplings from
 Eqs. (\ref{br-2}), (\ref{br-1}) and (\ref{br-3}) as
\bea
C_{10}^{NP}\Big[(0.58+0.128~ C_{10}+0.596~ C_7) \cos \phi^{NP}+0.02 \sin \phi^{NP}\Big]
&+&0.13 ~|C_{10}^{NP}|^2 = -0.32 \pm 0.51\nn\\&&({\rm  for~low}~q^2)\nn\\
C_{10}^{NP}\Big[(0.11 +0.03 ~C_{10}+0.07~C_7) \cos \phi^{NP}+0.009 \sin \phi^{NP}\Big]
&+&0.03 ~|C_{10}^{NP}|^2=0.06 \pm 0.12\nn\\ &&({\rm for ~ high}~q^2)
\eea

The corresponding 1-$\sigma$ allowed region in the $|C_{10}^{NP}|$ - $\phi^{NP}$ plane is shown in
the Figure-2 where the green region corresponds to the constraint coming from high-$q^2$ bound and the magenta region
coming from the low-$q^2$ limit. From the figure one can see that the bounds coming from the high-$q^2$
measurement is rather weak. From the  low-$q^2$ constraint one can infer that for the value $-1 \leq C_{10}^{NP}
\leq 1$ the entire range of $\phi^{NP}$ is allowed. These bounds can be
translated to the bounds on $r$ and $\phi^{NP}$ as done for $B_s \to \mu^+ \mu^-$ process as
\bea
0\leq r \leq 0.24\;,~~~~~~~~{\rm for}~~~~~~~~ 0 \leq \phi^{NP} \leq 2 \pi\;.\label{bn2}
\eea
Thus, from Eqns. (\ref{r-bnd}) and  (\ref{bn2}) one can see that the bounds on NP couplings coming  from ${\rm BR}(\bar B_d^0 \to X_s \mu^+ \mu^-)$ is slightly weak
in comparison to ${\rm BR}( B_s \to \mu^+ \mu^-)$.

\begin{figure}[htb]
\centerline{\epsfysize 3.0 truein \epsfbox{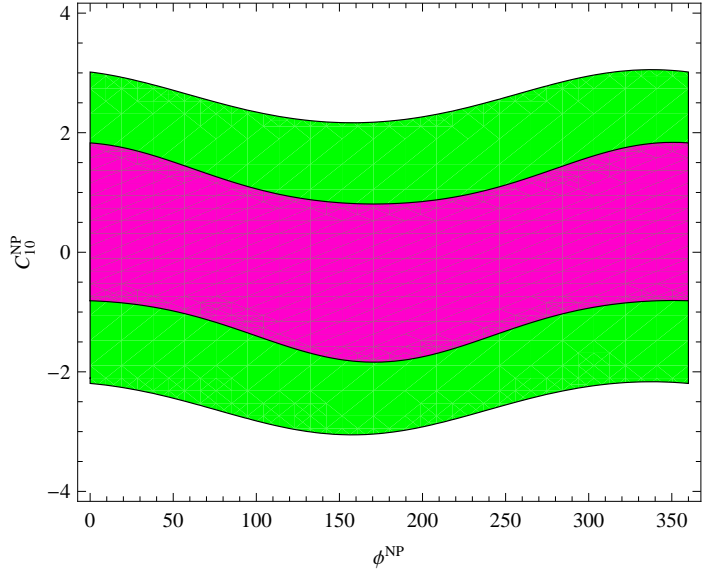}}
\caption{The allowed region in the $C_{10}^{NP}-\phi^{NP}$ parameters space obtained from the
${\rm BR}(\bar B_d \to X_s \mu^+ \mu^-)$, where the green (magenta) region corresponds to high-$q^2$
(low-$q^2$) limits.}
\end{figure}
Next we will consider the contributions coming from the $X(3,2,1/6)$ exchange. In this case  the new couplings $C_9^{'NP}$ and $C_{10}^{'NP}$ will come into picture. Proceeding in a similar fashion as done for $X(3,2,7/6)$ leptoquark case, we obtain
the constraint equations for these parameters as
\bea
0.064\Big[|C_9^{'NP}|^2 +|C_{10}^{'NP}|^2\Big]=(-0.32 \pm 0.51)~~~~~({\rm low}~q^2)\nn\\
0.014\Big[|C_9^{'NP}|^2 +C_{10}^{'NP}|^2\Big]=(0.06 \pm 0.12)~~~~~({\rm high}~q^2)\;.\label{bn-3}
\eea
The corresponding allowed region in $C_9^{'NP}$-$C_{10}^{'NP}$ plane is shown in Figure-3, where the green region corresponds
to the bounds coming from  high-$q^2$ limit and magenta region corresponds to the low-$q^2$ bound. Thus, from the low-$q^2$ bounds one can obtain  the limits on $C_9^{'NP}$ and $C_{10}^{'NP}$   as $-1.5 \leq |C_{9}^{'NP}|,|C_{10}^{'NP}| \leq 1.5$. Again translating the above bounds into the bound on $r$ one can obtain
\bea
0\leq r \leq 0.36\;,
\eea
which is again much weaker than the bounds coming from
$B_s \to \mu^+ \mu^-$ measurements.
\begin{figure}[htb]
\centerline{\epsfysize 3.0 truein \epsfbox{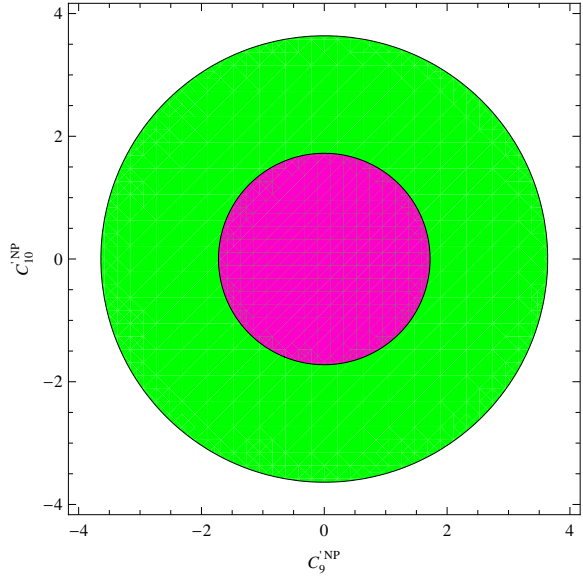}}
\caption{The allowed region in the $C_{9}^{'NP}-C_{10}^{'NP}$ parameters space obtained from the
${\rm BR}(\bar B_d \to X_s \mu^+ \mu^-)$, where the green (magenta) region corresponds to high-$q^2$
(low-$q^2$) limits.}
\end{figure}
However, in our analysis we will use relatively mild constraint, consistent
 with  both ${\rm BR}(B_s \to \mu^+ \mu^- )$ and ${\rm BR}(\bar B_d^0 \to X_s \mu^+ \mu^-)$ measurements as
 \bea
 0\leq r \leq 0.35\;, ~~~~{\rm with}~~~~60^\circ \leq \phi^{NP} \leq 270^\circ\;.\label{r-bound}
 \eea
This limit on $r$ can be translated to give us bound
on leptoquark coupling using Eqs. (\ref{c10np}), (\ref{c10np1}) and (\ref{bound}) as
\be
\left|\frac{\lambda_{\mu}^{23} \lambda_\mu^{22*}}{M_Y^2}\right|=\left|\frac{\lambda_{s}^{22} \lambda_b^{32*}}{M_V^2}\right|\leq 4.8 \times 10^{-9} ~{\rm GeV}^{-2}\;.
\ee
If we use the values of the couplings as $|\lambda_{d,e}|\approx 0.1$, allowing the perturbation theory to be valid,
we get the lower bound on the scalar leptoquark mass as
\be
M_{X}\geq 1.4 ~{\rm TeV}\;.
\ee

It should be noted that the recent measurement by LHCb collaboration \cite{lhcb13} shows several significant deviations
on angular observables in the rare decay $B \to K^{* 0} \mu^+ \mu^-$ from their corresponding SM expectations. In particular
an anomalously low value of $S_4$ at high $q^2$ at $2.8\sigma$ level and an opposite sign of $S_5$  at low $q^2$ region
at $2.4\sigma$ level. Although it is conceivable that these anomalies are due to statistical fluctuations or under estimated
theory uncertainties \cite{ref28}, but the possible indication of new physics could not be ruled out. It has been shown in Ref.
\cite{ref29} that a consistent explanation of most of the anomalies associated with $b \to s$ rare decays can be obtained by
NP contributing simultaneously to the semileptonic operator $O_9$ and its chirally flipped counterpart $O_9'$ with
$C_9^{NP} \simeq -(1.0 \pm 0.3)$ and $C_9^{'NP} \simeq 1.0 \pm 0.5$. However, in the leptoquark model since $C_9^{NP}$ and $C_{10}^{NP}$ contribute simultaneously it may not be possible to explain these anomalies.

After obtaining the allowed range for the leptoquark coupling we will now proceed to study the semileptonic
decay process $B_s \to \phi \mu^+ \mu^-$.

\section{$B_s \to \phi~ l^+ l^- $ process}

Here we will consider the decay mode $B_s \to \phi \mu^+ \mu^-$.
At the quark level, this decay mode  proceeds
through the FCNC transition $ b\to s l^+ l^-$, which occurs only through
loops in the SM, and therefore, it constitutes a quite
suitable tool of looking for new physics.
Moreover, the dileptons present in this process allow us to
formulate many observables which can serve as a testing ground to
decipher the presence of new physics \cite{ghosh}.

Recently the branching ratio of this decay mode
has been measured by the LHCb collaboration \cite{lhcb-phi} using the data corresponding to an integrated
luminosity of $1.0 ~{\rm fb}^{-1}$ collected at $\sqrt s$=7 TeV as
\be
{\rm BR}(B_s^0 \to \phi \mu^+ \mu^-) = \left (7.07_{-0.59}^{+0.64} \pm 0.17 \pm 0.71 \right ) \times 10^{-7}\;.
\label{phi-br}
\ee
They have also performed the angular analysis and determine the angular observables $F_L$, $S_3$,
$A_6$ and $A_9$, which are consistent with the standard model expectations.
This process has been very well studied in the
literature, both in the SM and in various extensions of it \cite{phi2}.
The branching ratio predicted in the standard model is in the range  $(14.5 - 19.2) \times 10^{-7}$ which is
significantly higher than the present experimental value (\ref{phi-br}). This deviation may be considered as
a smoking gun signal of new  physics in this mode or more generally in the processes involving
$b \to s$ transitions.

Using the effective Hamiltonian presented in Eq. (1) one can obtain the transition amplitude for this process.
The matrix elements of the various hadronic currents between the initial $B_s$ meson and the final vector meson
$\phi$ can be parameterized in terms of various form factors as \cite{ball}
\bea
&&\langle \phi(k, \varepsilon)|(V-A)_\mu |B_s(P) \rangle =
 \epsilon_{\mu \nu \alpha \beta} \varepsilon^{* \nu} P^\alpha k^\beta
\frac{2 V(q^2)}{m_{B}+m_\phi}-i \varepsilon_\mu^*(m_{B}+m_\phi)
A_1(q^2)\nn\\
&&~~~~~~~~~~~~+
i(P+k)_\mu (\varepsilon^* q)\frac{A_2(q^2)}{m_{B}+m_\phi}
+i q_\mu(\varepsilon^*  q) \frac{2 m_\phi}{q^2}\Big[A_3(q^2)-A_0(q^2)
\Big]\;,
\nn\\ \nn\\
&&\langle \phi (k, \varepsilon)|\bar s \sigma_{\mu \nu} {q^\nu}(1 +
\gamma_5) b |B_s(P) \rangle  =
i \epsilon_{\mu \nu \alpha \beta} \varepsilon^{* \nu} P^\alpha k^\beta
2 T_1(q^2)+ \Big[\varepsilon_\mu^*(m_{B}^2-m_\phi^2)\nn\\
&&~~~~~~~~
 -
(\varepsilon^*  q)(P+k)_\mu\Big]T_2(q^2)
+  (\varepsilon^*  q)\Big[q_\mu -
\frac{q^2}{m_{B}^2-m_\phi^2}(P+k)_\mu \Big]T_3(q^2)\;,\label{vf}
\eea
where $V$ and $A$ denote the vector and axial vector currents,
$A_0,A_1,A_2,A_3,V,T_1,T_2$ and $T_3$ are the relevant form factors
and $q$ is the momentum transfer.

Thus, with eqs. (\ref{ham}) and (\ref{vf})
the transition amplitude for $B_s \to \phi l^+ l^-$ is given as
\bea
{\cal M}(B_s  \to  \phi~ l^+ l^-) & = & \frac{G_F \alpha}
{2 \sqrt 2 \pi} V_{tb} V_{ts}^* \biggr\{ \bar l \gamma^\mu l \Big[
-2 A \epsilon_{\mu \nu \alpha \beta}\varepsilon^{* \nu} k^\alpha q^\beta
-iB \varepsilon_\mu^*\nn\\
&+ & i C (P+k)_\mu (\varepsilon^*\cdot q)
+  iD (\varepsilon^* \cdot q)
q_\mu \Big]
+ \bar l \gamma^\mu \gamma_5 l \Big[- 2E \epsilon_{\mu \nu \alpha \beta}
\varepsilon^{* \nu} k^\alpha q^\beta \nn\\
&- &iF \varepsilon_\mu^*
+ i G  (\varepsilon^* \cdot  q)(P+k)_\mu +iH (\varepsilon^* \cdot q)
q_\mu \Big]\biggr\},\label{phi}
\eea

where the parameters $A,B, \cdots H$ are given as \cite{london}
\bea
A &=& 2 \left (C_9^{eff~SM} +C_9^{NP}+C_9^{'NP} \right )\frac{V(q^2)}{m_B + m_\phi}+
4 \frac{m_b}{q^2} C_7 T_1(q^2)\;, \nn\\
B &=& (m_B + m_\phi) \left (
2( C_9^{eff~SM}+C_9^{NP} -C_9^{'NP}) A_1(q^2) + 4 \frac{m_b}{q^2}(m_B -m_\phi) C_7 T_2(q^2)
\right )\;,\nn\\
C &=& 2( C_9^{eff~SM}+C_9^{NP} -C_9^{'NP}) \frac{A_2(q^2)}{m_B+m_\phi} +4 \frac{m_b}{q^2} C_7
\left (T_2(q^2) + \frac{q^2}{m_B^2-m_\phi^2} T_3(q^2) \right )\;,\nn\\
D &=& 4( C_9^{eff~SM}+C_9^{NP} -C_9^{'NP})\frac{m_\phi}{q^2}\Big(A_3(q^2)-A_0(q^2)
\Big)-4 C_7 \frac{m_b}
{q^2}T_3(q^2) \;,\nn\\
E&=& (C_{10}^{SM}+C_{10}^{NP}+C_{10}^{' NP}) \frac{2V(q^2)}{m_B+m_\phi}\;,\nn\\
F &=& 2(C_{10}^{SM}+C_{10}^{NP}-C_{10}^{' NP})(m_B +m_\phi) A_1(q^2)\;,\nn\\
G&=& (C_{10}^{SM}+C_{10}^{NP}-C_{10}^{' NP}) \frac{2A_2(q^2)}{m_B+m_\phi}\;,\nn\\
H &=& 4 (C_{10}^{SM}+C_{10}^{NP}-C_{10}^{'NP}) \frac{m_\phi}{q^2}\Big(A_3(q^2)-A_0(q^2)\Big)\;.
\eea
The differential decay rate is given as
\be
\frac{d \Gamma}{d s}=\frac{G_F^2 \alpha^2}{2^{14} \pi^5}
|V_{tb}V_{ts}^*|^2 m_B \tau_B
\lambda^{1/2} (1, r_\phi, \hat s)~ v_l ~ \Delta\;,\label{lp1}
\ee
where $\hat s=q^2/m_B^2$, $r_\phi=m_\phi^2/m_B^2$, $v_l=
\sqrt{1-4m_l^2/s}$,
$\lambda \equiv \lambda(1,r_\phi,\hat s)$, is the triangle function and
\bea
\Delta &=& \frac{1}{3 r_\phi} \Big [ 8 \lambda m_B^4 \hat s\Big((3-v_l^2)|A|^2+
(12 r_\phi \hat s+\lambda)(3-v_l^2) |B|^2\nn\\
&+& \lambda^2 m_B^4 (3-v_l^2)|C|^2
+16 v_l^2 m_B^4 r_\phi \hat s \lambda |E|^2 +
(24 r_\phi \hat s v_l^2 +\lambda (3-v^2))|F|^2 \nn\\
&+& m_B^4 \lambda \left ( 6 \hat s (1+r_\phi) (1-v^2) -
3 \hat s^2(1-v_l^2) + \lambda (3-v^2)\right )|G|^2\nn\\
&+& 3 \lambda m_B^4 \hat{s}^2 (1-v_l^2)|H|^2 + 2 Re [F G^*]m_B^2 \lambda
\left (r_\phi(3-v^2) +v_l^2(1+ 2\hat s)-3 \right )\nn\\
&-&6 Re[F H^*] m_B^2 \hat s (1-v_l^2) \lambda + 6 Re[G H^*]m_B^4 \hat s
\lambda (1-r_\phi)(1-v^2)\nn\\
&+&2 Re[B C^*] m_B^2 \lambda (3-v^2)(r_\phi +\hat s -1) \Big]
\;.
\eea
Another observable is the lepton forward backward asymmetry ($A_{FB}$),
which is also a very powerful tool for looking into  new physics signature. In particular the
position of the zero value of $A_{FB}$ is very sensitive to
the presence of new physics. The normalized forward-backward asymmetry
is defined as
\bea
A_{FB}(s) = \frac{\ds{\int_0^1 \frac{d^2 \Gamma}{d \s1 d \cos \theta}
d \cos \theta-\int_{-1}^0
\frac{d^2 \Gamma}{d \s1 d \cos \theta}d \cos \theta}}
{\ds{\int_0^1 \frac{d^2 \Gamma}{d \s1 d \cos \theta}d
\cos \theta +\int_{-1}^0
\frac{d^2 \Gamma}{d \s1 d \cos \theta} d \cos \theta}}\;,\label{fb}
\eea
where $\theta $ is the angle between the directions of
$l^+$ and $B_s$ in the rest frame of the lepton pair.
The forward-backward asymmetry can also be written in the form \cite{london}
\bea
A_{FB}(q^2) = -\frac{1}{\Delta}8 m_B^2 ~\sqrt{\lambda}~ v_l~ \hat s ~Re[A^* F + B^* E]
\eea
As seen from \cite{lhcb-phi}, the actual decay being observed is not $B_s \to \phi \mu^+ \mu^-$ but
$B_s \to \phi(\to K^+ K^-) \mu^+ \mu^-$. Thus, the angular analysis of the four body final state
offers a large number of observables in the differential decay distribution \cite{phi3}.
The angular distribution of the decay process ${\bar B}_s^0 \to \phi (\to K^+ K^-) \mu^+ \mu^-$ can be defined by the
decay angles $\theta_K$, $\theta_l$ and $\Phi$, where $\theta_K~(\theta_l)$ denotes the angle
of $K^-~(\mu^-)$ with respect to the direction of flight of the $\bar{B}_s$ meson in the $K^+ K^-(\mu^+ \mu^-)$
center-of-mass frame and $\Phi$ denotes relative angle of the $\mu^+ \mu^-$ and the $K^+ K^-$ decay planes
in the $\bar{B}_s$ meson center of mass frame and is given as \cite{buras2} as
\bea
\frac{d^4 \Gamma}{d q^2~ d \cos \theta_l ~d \cos \theta_K ~d \Phi} &=& \frac{9}{32 \pi} I_1^s \sin^2 \theta_K + I_1^c
\cos^2 \theta_K +(I_2^s \sin^2 \theta_K +I_2^c \cos^2 \theta_K) \cos 2 \theta_l \nn\\
&+& I_3 \sin^2 \theta_K \sin^2 \theta_l \cos 2 \Phi +I_4 \sin 2 \theta_K \sin 2 \theta_l \cos \Phi \nn\\
&+& I_5 \sin 2 \theta_K \sin \theta_l \cos \Phi +(I_6^s \sin^2 \theta_K+I_6^c \cos^2 \theta_K ) \cos \theta_l \nn\\
&+& I_7 \sin 2 \theta_K \sin \theta_l \sin \Phi +I_8 \sin 2 \theta_K \sin 2 \theta_l \sin \phi \nn\\
&+& I_9 \sin^2 \theta_K \sin^2 \theta_l \sin 2 \Phi\;.\label{angular}
\eea
The corresponding expression for CP conjugate process  $B_s^0 \to \phi (\to K^+ K^-) \mu^+ \mu^-$  ($d^4 \bar{\Gamma}$) can be obtained from
(\ref{angular}) by the replacement of   $I_i$'s  by ${\bar I_i}$'s where these observables are related to each other through
\be
 I_{1,2,3,4,7}^{(a)} \longrightarrow \bar{I}_{1,2,3,4,7}^{(a)}\;,~~~~~~~~~~~~~  I_{5,6,8,9}^{(a)} \longrightarrow -\bar{I}_{5,6,8,9}^{(a)}\;,
\ee
with all weak phases conjugated. The angular coefficients $I_i^{(a)}$, usually expressed in terms of the transversity  amplitudes
which are given as \cite{buras2}
\bea
A_{\perp L,R}&=& N\sqrt{2  \lambda_1}\Big [ \left ( (C_9^{eff}+C_9^{NP} +C_9^{'NP}) \mp (C_{10}+C_{10}^{NP} +C_{10}^{'NP} ) \right )
\frac{V(q^2)}{m_B+m_{\phi}} + {2 m_b}{s} C_7 T_1(q^2) \Big]\;,\nn\\
A_{\para L,R}&=& -N \sqrt{2} (m_B^2 -m_{\phi}^2)\Big[ \left ((C_9^{eff}+C_9^{NP}-C_9^{'NP}) \mp (C_{10}+C_{10}^{NP}-C_{10}^{'NP} ) \right ) \frac{A_1(q^2)}{m_B - m_\phi}\nn\\
&& + \frac{2 m_b}{s}C_7 T_2(q^2) \Big]\;,\nn\\
A_{0 L,R}& =& - \frac{N}{2 m_{\phi} \sqrt{s}} \Big[ \left (C_9^{eff}+C_9^{NP}-C_9^{'NP}) \mp (C_{10}+C_{10}^{NP}-C_{10}^{'NP}) \right ) \nn\\
&& \times \left ( (m_B^2 -m_\phi^2 -s)(m_B + m_\phi) A_1(q^2)-\lambda_1 \frac{A_2(q^2)}{m_B + m_\phi} \right )\nn\\
&&+ 2 m_B C_7 \left ( (m_B^2 +3 m_\phi^2 -s) T_2(q^2) - \frac{\lambda_1}{m_B^2 - m_\phi^2} \right ) \Big]\;,\\
A_t&=& N \frac{\lambda_1}{s} \Big[ 2 (C_{10}+C_{10}^{NP}-C_{10}^{'NP}) \Big] A_0(q^2)
\eea
where
\be
N= V_{tb}V_{ts}^* \left [ \frac{G_F^2 \alpha^2}{3 \cdot 2^{10} \pi^5 m_B^3} s v \sqrt{\lambda_1 } \right ]^{1/2}\;,
\ee
with $\lambda_1= (m_B^2 + m_\phi^2 +s)^2-4 m_B^2 m_\phi^2 $.
With these transversity amplitudes the angular coefficients are given as
\bea
I_1^s &=& \frac{2+v_l^2}{4} \Big[ | A_\perp^L|^2 + |A_\para^L|^2  + (L \to R) \Big]+
\frac{4 m_\mu^2}{s} Re \left (A_\perp^L A_\perp^{R *} + A_\para^L A_\para^{R*} \right )\nn\\
I_1^c &=& |A_0^L|^2 +|A_0^R|^2 + \frac{4 m_\mu^2}{s} \Big( |A_t|^2 + 2 Re (A_0^L A_0^{R *}) \Big)\nn\\
I_2^s & = & \frac{v_l^2}{4} \Big( |A_{\perp}^L|^2 +|A_\para^L|^2 + (L \to R) \Big)\;,\nn\\
I_2^c &= & -v_l^2 \Big(|A_{0}^L|^2 + (L \to R)   \Big)\;,\nn\\
I_3 &=& \frac{v_l^2}{2} \Big( Re(A_0^L A_\para^{L *} + (L \to R) \Big)\;,\nn\\
I_4 &=& \frac{v_l^2}{2} \Big(Re(A_0^L A_\para^{L*} -(L \to R) \Big)\;,\nn\\
I_5 &=& \sqrt{2} v_l \Big( Re(A_\para^L A_\perp^{L *} - (L \to R)\Big)\;, \nn\\
I_6^s &=& 2 v \Big( Re(A_\para^L A_\perp^{L *}- (L \to R) \Big) \;, \nn\\
I_7 &=& \sqrt{2} v_l \Big ( Im(A_0^L A_\para^{L*}- (L \to R) \Big)\;,\nn\\
I_8& =& \frac{v_l^2}{\sqrt 2} \Big( Im(A_0^L A_\perp^{L *})+(L \to R)\Big) \;,\nn\\
I_9 &=& v_l^2\Big( Im (A_\para^{L*} A_\perp^L) + (L \to R)\Big)\;.
\eea
From these angular coefficients one can construct twelve CP averaged angular coefficients $S_i^{(a)}$ and twelve CP asymmetries
$A_i^{(a)}$ as
\bea
S_i^{(a)} = (I_i^{(a)}+\bar{I}_i^{(a)})\Big/\frac{d (\Gamma + \bar{\Gamma})}{d q^2}\;,
~~~~~~~A_i^{(a)} = (I_i^{(a)}-\bar{I}_i^{(a)})\Big/\frac{d (\Gamma + \bar{\Gamma})}{d q^2}\;.
\eea
All the physical observables can be expressed in terms of $S_i$ and $A_i$. For example the CP asymmetry
in the dilepton mass distribution can be expressed as
\bea
A_{CP}= \frac{d (\Gamma - \bar{\Gamma})}{d q^2}\Big/\frac{d (\Gamma + \bar{\Gamma})}{d q^2}= \frac{3}{4}
(2 A_1^s +A_1^c) - \frac{1}{4}(2 A_2^s +A_2^c)\;.
\eea
The $q^2$ average of these observables are defined as follows:
\bea
\langle S_i^{(a)}\rangle & = &\int_{1~{\rm GeV^2}}^{6~{\rm GeV^2}} d q^2 (I_i^{(a)}+\bar{I}_i^{(a)})\Bigg/\int_{1~{\rm GeV^2}}^{6~{\rm GeV^2}} d q^2
\frac{d (\Gamma + \bar{\Gamma})}{d q^2}\nn\\
\langle A_i^{(a)}\rangle & = &\int_{1~{\rm GeV^2}}^{6~{\rm GeV^2}} d q^2 (I_i^{(a)}-\bar{I}_i^{(a)})\Bigg/\int_{1~{\rm GeV^2}}^{6~{\rm GeV^2}} d q^2
\frac{d (\Gamma + \bar{\Gamma})}{d q^2}\;.
\eea
 After getting familiar with the different observables associated with $B_s \to \phi \mu^+ \mu^-$ decay process we now proceed for numerical
estimation. For this purpose we use the form factors calculated in the light-cone sum rule (LCSR)
approach \cite{ball}, where the $q^2$ dependence of various form factors are given
by simple fits as
\bea
f(q^2) &=& \frac{r_2}{1-q^2/m_{fit}^2}\;,~~~
~~~~~~~~~~~~~~~~~~~~~~~~~~({\rm for}~~A_1,~T_2)\nn\\
f(q^2) &= & \frac{r_1}{1-q^2/m_{R}^2}+
\frac{r_2}{1-q^2/m_{fit}^2}\;,
~~~~~~~~~~({\rm for}~~V,~ A_0,~ T_1)\nn\\
f(q^2) &= & \frac{r_1}{1-q^2/m_{fit}^2}+  \frac{r_2}{(1-q^2
/m_{fit}^2)^2}\;,
~~~~~~({\rm for}~~A_2, ~\tilde T_3)\;.
\eea
The values of the parameters $r_1$, $r_2$, $m_R$ and  $m_{fit}$
are taken from \cite{ball}. The form
factors $A_3$ and $T_3$ are given as
\bea
A_3(q^2) &=& \frac{m_B+m_V}{2 m_\phi}A_1(q^2) -
\frac{m_B-m_\phi}{2 m_\phi} A_2(q^2)\;,\nn\\
T_3(q^2) &=& \frac{m_B^2-m_\phi^2}{q^2}\Big(\tilde{ T}_3(q^2)-T_2(q^2)\Big).
\eea
The particle masses and the lifetime of $B_{s}$ meson are taken from \cite{pdg}. The quark masses
(in GeV) used are
$m_b$=4.8, $m_c$=1.5, the fine structure coupling constant $\alpha=1/128$   and the CKM matrix elements as
$V_{tb} V_{ts}^*=0.0405$. Using these values we show in Figure-4 the variation of differential decay rate (left panel) and the forward
backward asymmetry (right panel)  in the standard model with respect to the di-muon invariant mass.

\begin{figure}[htb]
\includegraphics[width=7.5cm,height=5.0cm, clip]{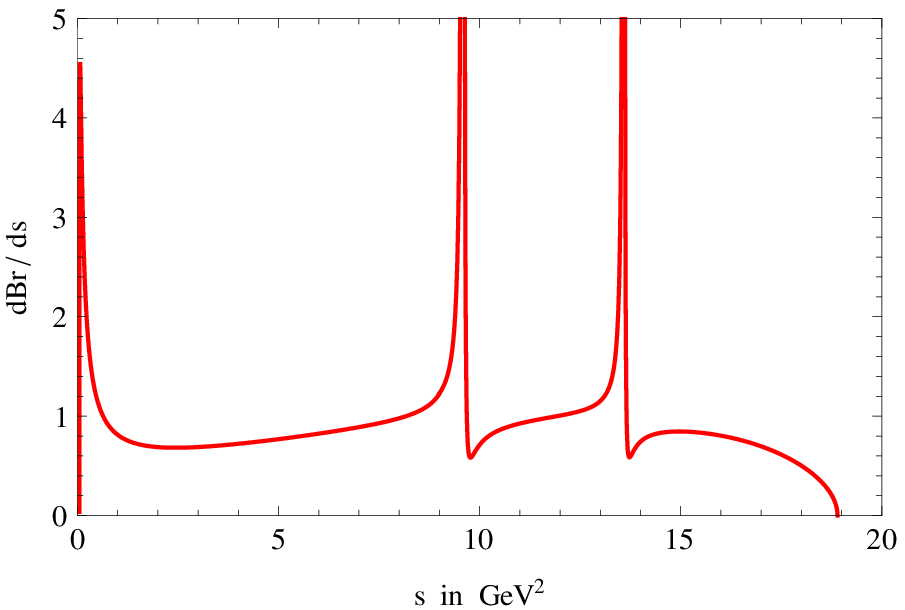}
\hspace{0.2 cm}
\includegraphics[width=7.5cm,height=5.0cm, clip]{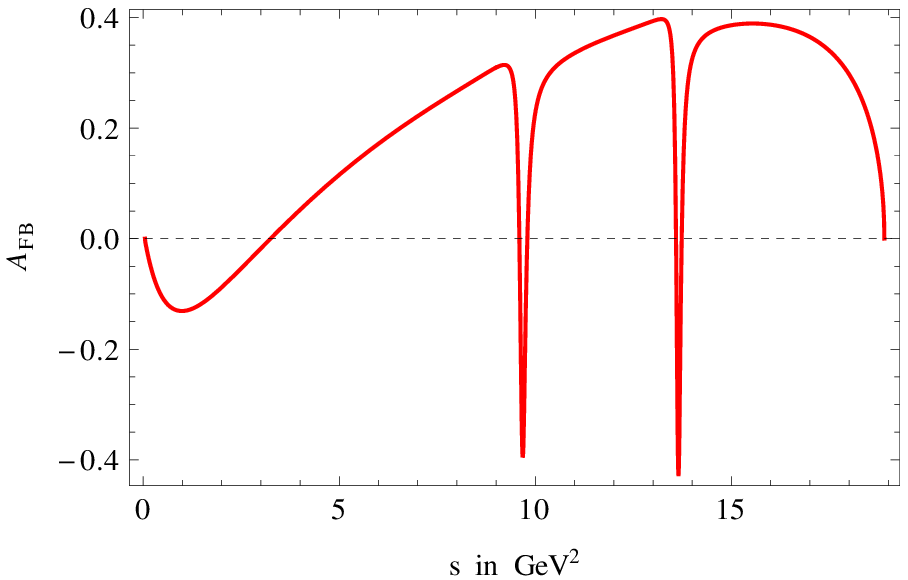}
\caption{Variation of the differential branching ratio (in units of $10^{-7}$)
(left panel) and the forward-backward  asymmetry with respect to the
momentum transfer $s$  (right panel) for
the $B_s \to \phi \mu^+ \mu^-  $  process. }
\end{figure}

In the leptoquark model, this process will receive additional contribution arising from the leptoquark
exchange. Hence, in the leptoquark model the Wilson coefficients $C_{9,10}$ will receive additional contributions
$C_{9,10}^{NP}$  as well as
new Wilson $C_{9,10}'$ associated with the chirally flipped operators $O_{9,10}'$ will also be present
as already discussed in Section III. The bounds on these new coefficients can be obtained from
the constraint on $r$ (\ref{r-bound}) extracted from the
experimental results on BR$(B_s \to \mu^+ \mu^-)$ and BR$(\bar B_d^0 \to X_s \mu^+\mu^-)$. For the leptoquarks $X=(3,2,7/6)$ and $X=(3,2,1/6)$,
we obtain the value of $r\leq 0.35$ for $\phi$ in the range $(60-270)^\circ $. This constraint can be
translated with eqns (9), (13) and (\ref{r-bound}) which gives the value of the new Wilson coefficients as
\bea
&&|C_9^{LQ}|= |C_{10}^{LQ}|\leq |r ~C_{10}^{SM}| ~~~~~~~~({\rm for}~X=(3,2,7/6))\nn\\
&&|C_9^{'~LQ}|=|C_{10}^{'~LQ}| \leq |r~C_{10}^{SM}|~~~~~~({\rm for}~X=(3,2,1/6))\;.
\eea

Using these values we show the variation of differential decay rate and forward-backward asymmetry
for $X=(3,2,7/6)$ in Figure-5 and for $X=(3,2,1/6)$  in Figure-6. From these figures it can be
seen that the branching ratio could have significant deviation from its SM value both in the upward
as well as downward direction. However, the zero position of the forward-backward asymmetry does not
have any significant deviation.
\begin{figure}[htb]
\includegraphics[width=7.5cm,height=5.5cm, clip]{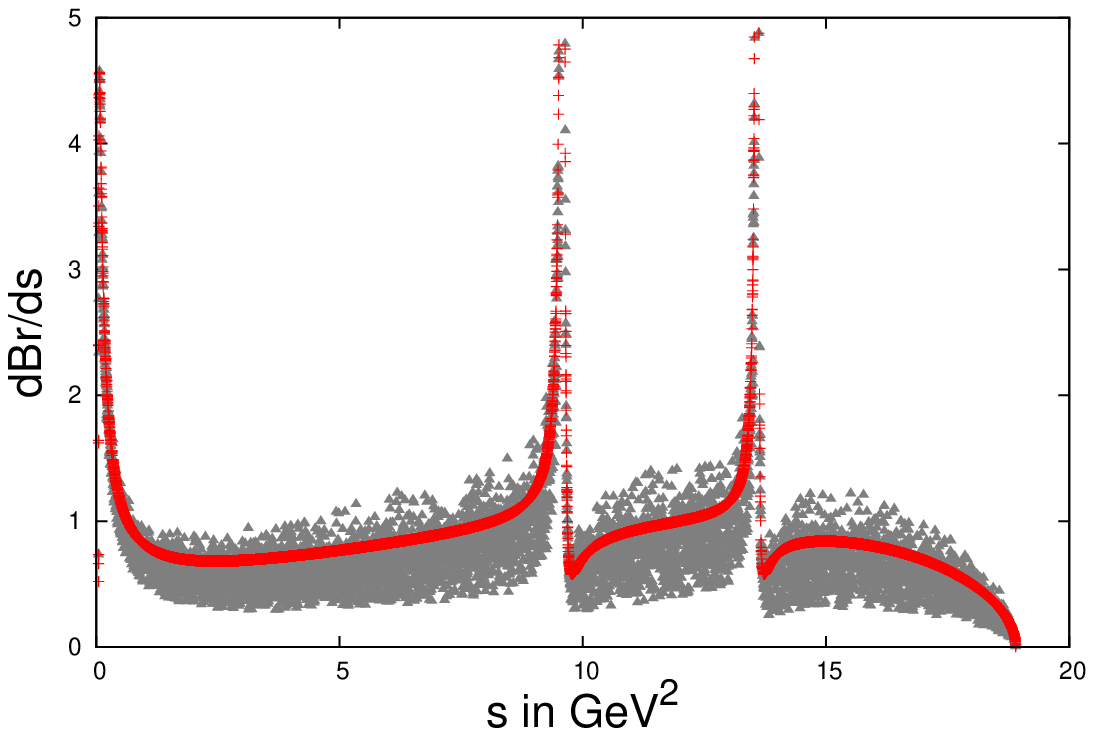}
\hspace{0.2 cm}
\includegraphics[width=7.5cm,height=5.5cm, clip]{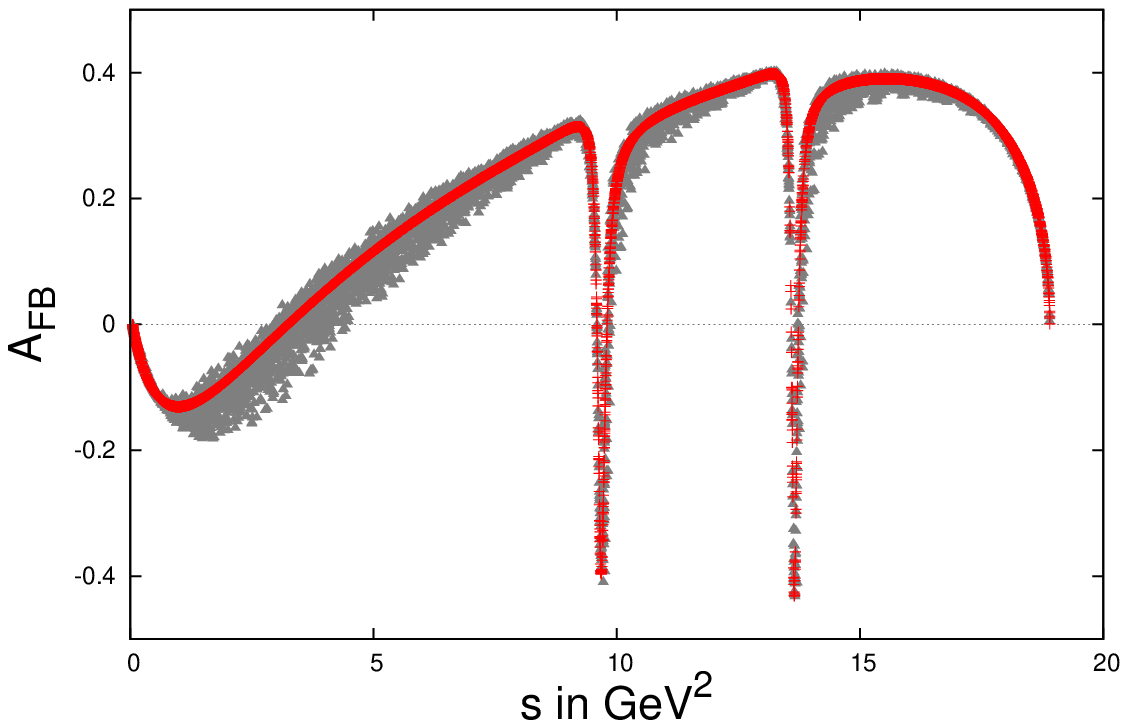}
\caption{Same as Figure-4, where the red curves represent the SM values and the grey regions represent
the results due to $X=(3,2,7/6)$ leptoquark contributions.}
\end{figure}

\begin{figure}[htb]
\includegraphics[width=7.5cm,height=5.5cm, clip]{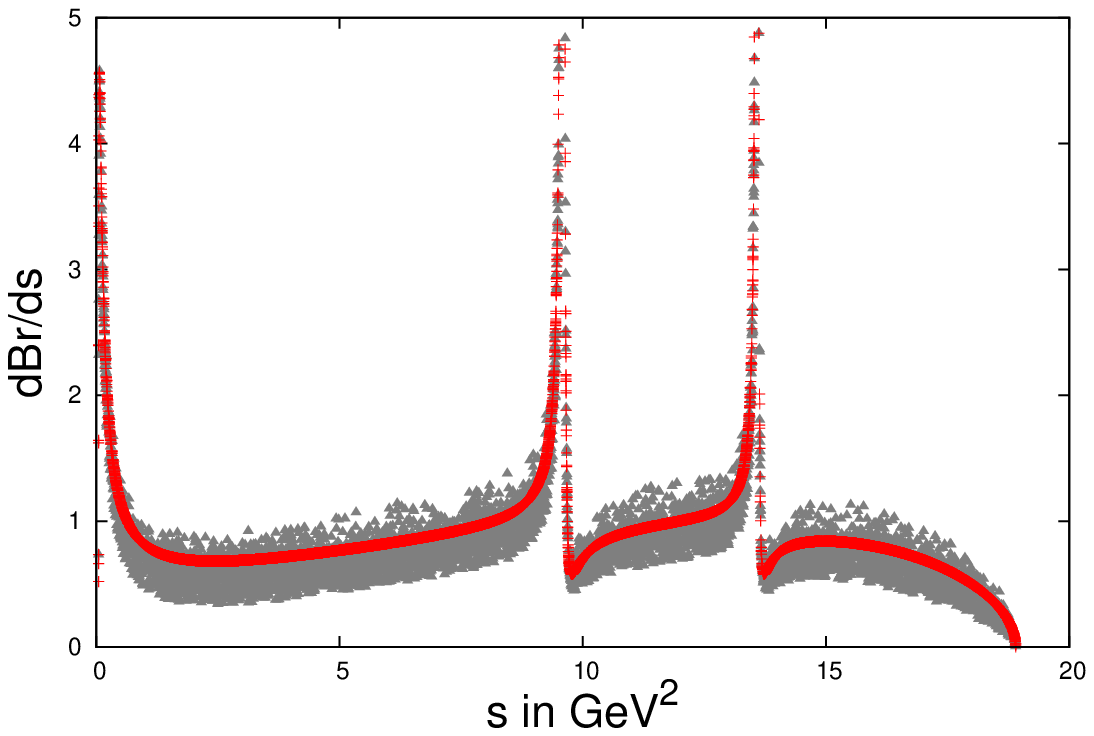}
\hspace{0.2 cm}
\includegraphics[width=7.5cm,height=5.5cm, clip]{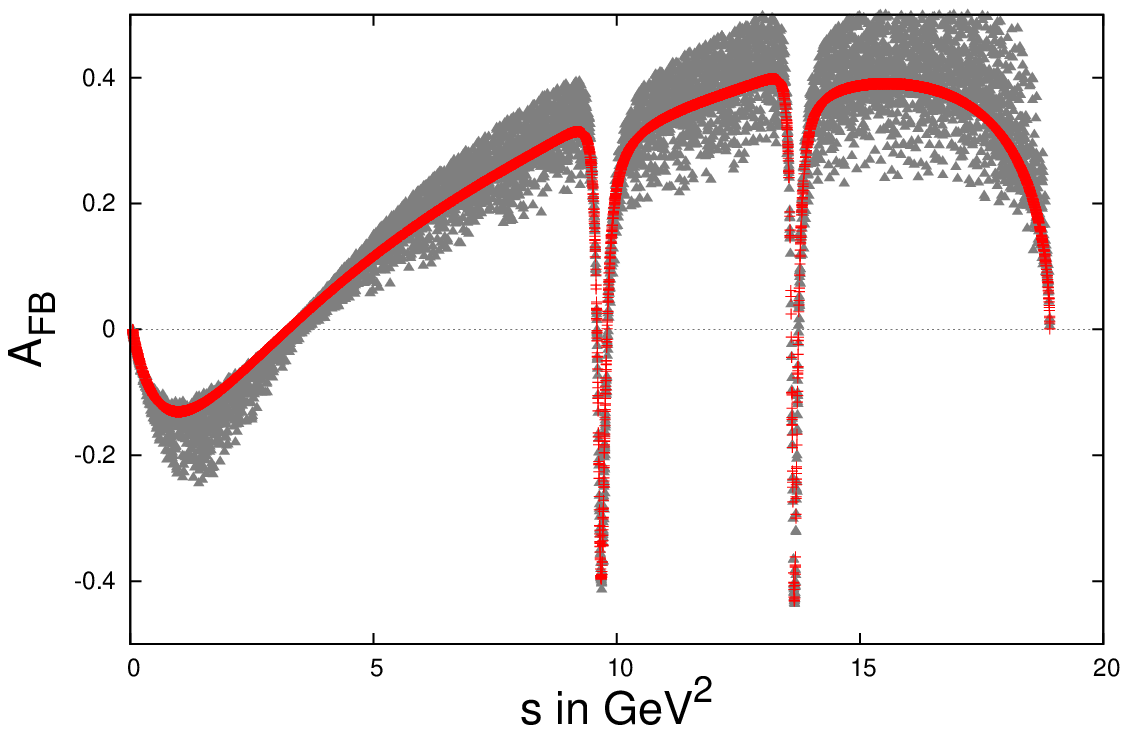}
\caption{Same as Figure-4, where the red curves represent the SM values and the grey regions represent
the results due to $X=(3,2,1/6)$ leptoquark contributions.}
\end{figure}
We now proceed to calculate the  total decay rate for $B_s \to
\phi~ \mu^+ \mu^-$. It should be noted that the long distance
contributions arise from the real $\bar c c$ resonances with the
dominant contributions coming from the low
lying resonances $J/\psi$ and $\psi'(2S)$. In order to
minimize the hadronic uncertainties it is necessary to eliminate
the backgrounds coming from the resonance regions.
The resonant decays $B_s \to J/\psi \phi$ and $B_s^0 \to \psi'(2S) \phi$ with $\psi/\psi'(2S) \to \mu^+ \mu^-$ are
rejected by applying the vetos on the dimuon mass regions around the charmonium resonances, i.e.,
$(2946 < m(\mu^+ \mu^-) < 3176)~{\rm MeV/c^2}$ and $(3592 < m(\mu^+ \mu^-) < 3766)~ {\rm MeV/c^2}$ \cite{lhcb-phi}.
Using these veto windows we obtain the branching ratio for
the  $B_s \to \phi \mu^+ \mu^-$ decay mode as
\bea
{\rm BR}(B_s \to \phi \mu^+ \mu^-) &= & 13.2 \times 10^{-7}~~~~~~~~~({\rm in~SM})\;, \nn\\
&=& (5.8 -24.4) \times 10^{-7}~~~~~({\rm in~ LQ ~Model-I~(X=3,2,7/6)})\;,\nn\\
&=& (8.1 -22.0) \times 10^{-7}~~~~~({\rm in~ LQ ~Model-II~(X=3,2,1/6)})\;.
\eea
Thus, one can see that the observed branching ratio (\ref{phi-br}) can be accommodated in the scalar leptoquark model.

Our next objective is to study the effect of leptoquark in the CP asymmetry parameters $A_i^{(a)}$.
The $q^2$ variation of these observables in the low $q^2$ regime is shown in Figure-7. Here  we have varied
new weak phase between (60-90) degree and fixed the $r$ value at 0.35. The time integrated value in the low $q^2$
region is shown in Table-1. Some of these asymmetries are measured by the LHCb collaboration, which are almost in
agreement with the standard model predictions but with large error bars. Future measurement with large data samples
could possibly minimize these errors and help to infer the presence of new physics, if there is any from these
observables.

\begin{table}
 \begin{center}
\caption{Allowed range of the CP violating observables (in units of $10^{-3})$, in the leptoquark model. }
  \begin{tabular}{c|c|c|c}

   \hline
Observables  & Allowed range & Observables & Allowed range \\
& (in units of $10^{-3}$) & &(in units of $10^{-3})$\\
\hline

$\langle A_1^s \rangle  $ & $(0.18 \to 0.27)$ &  $\langle A_5 \rangle  $ & $ -(60 \to 110)$\\

$\langle A_1^c \rangle  $ & $(8 \to 12)$ &  $\langle A_6^s \rangle  $ & $(7.6 \to 8.0)$\\

$\langle A_2^s \rangle  $ & $ (0.06 \to 0.09)$ &  $\langle A_8 \rangle  $ & $(3.8 \to 4.0) $\\

$\langle A_2^c \rangle  $ & $-(7.9 \to 41.8)$ &  $\langle A_7 \rangle  $ & $ -(46 \to 67) $\\

$\langle A_3 \rangle  $ & $-(1.3 \to 1.9)$ &  $\langle A_9 \rangle  $ & $-(0.12 \to 0.84) $\\

$\langle A_4 \rangle  $ & $(2.3 \to 3.4)$ &  $\langle  A_{CP}\rangle  $ & $(8.4 \to 12.0)$\\

\hline

\end{tabular}
\end{center}
\end{table}

\begin{figure}[htb]
\begin{center}
\includegraphics[width=5cm,height=3.5cm]{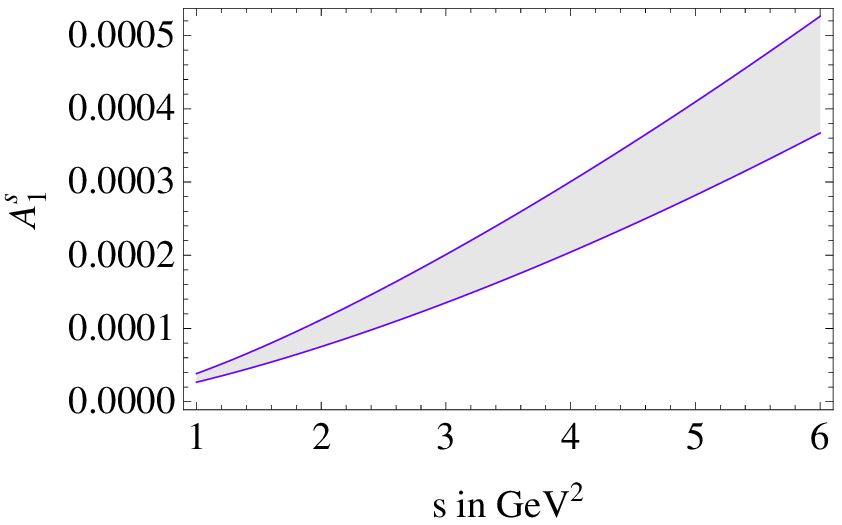}
\includegraphics[width=5cm,height=3.5cm]{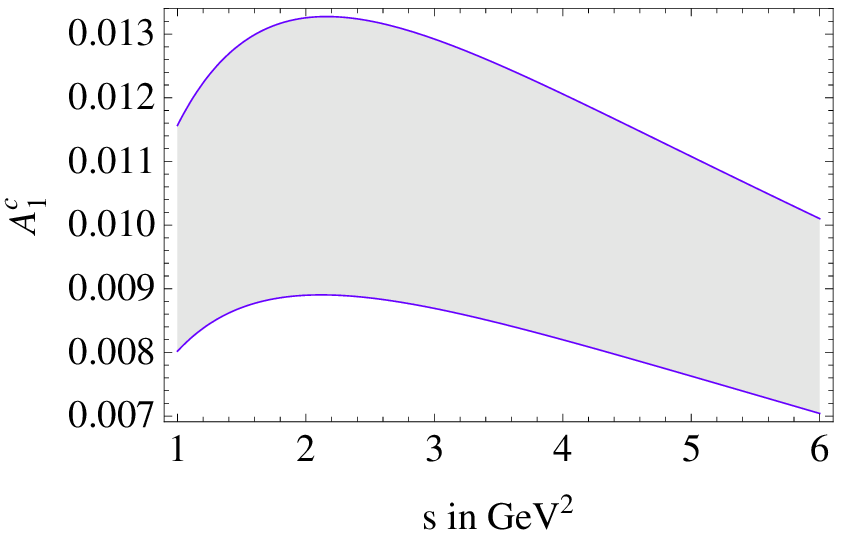}
\includegraphics[width=5cm,height=3.5cm]{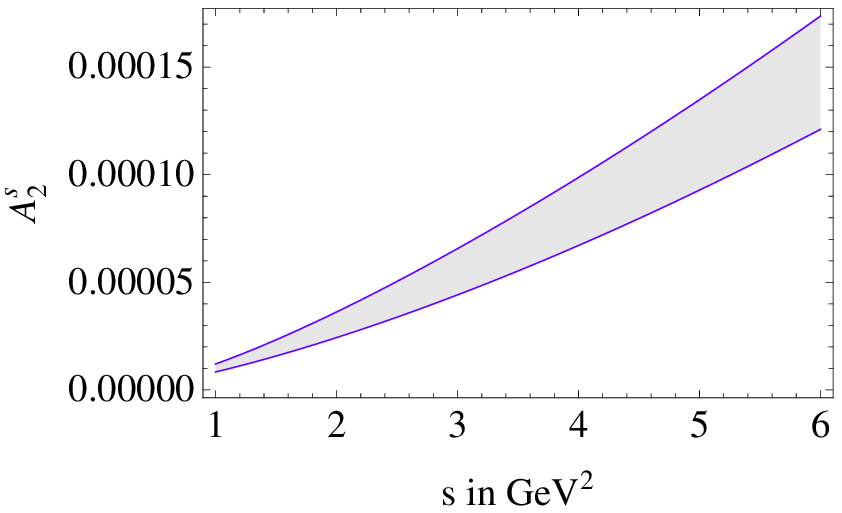}
\includegraphics[width=5cm,height=3.5cm]{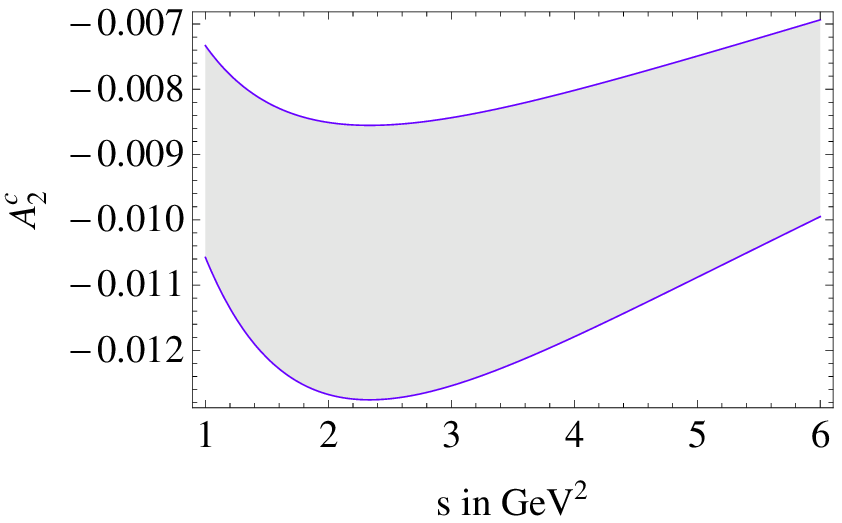}
\includegraphics[width=5cm,height=3.5cm]{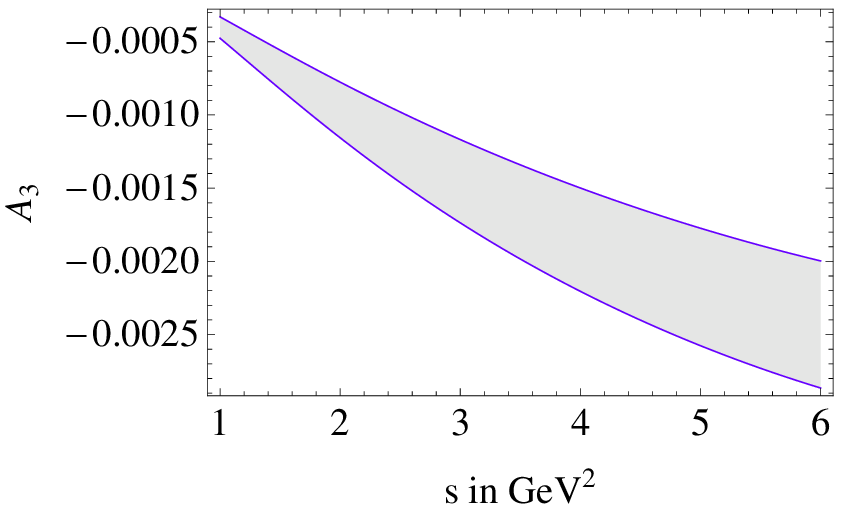}
\includegraphics[width=5cm,height=3.5cm]{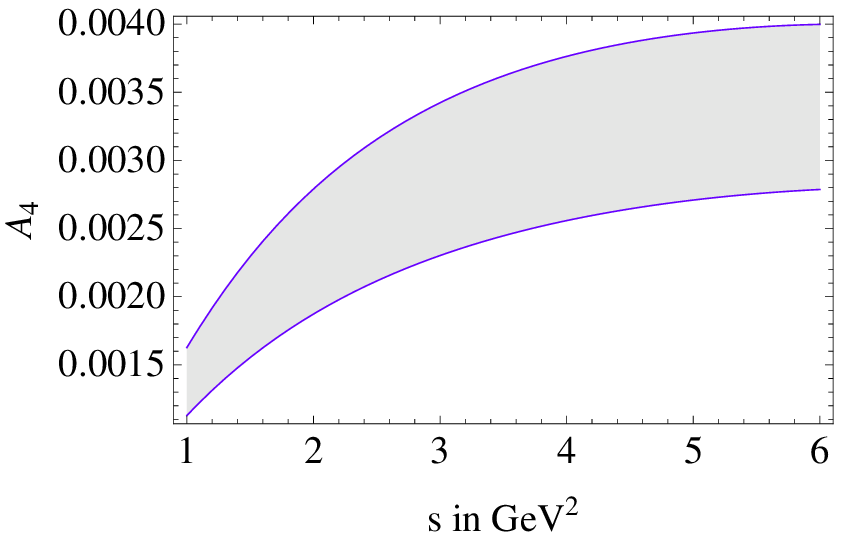}
\includegraphics[width=5cm,height=3.5cm]{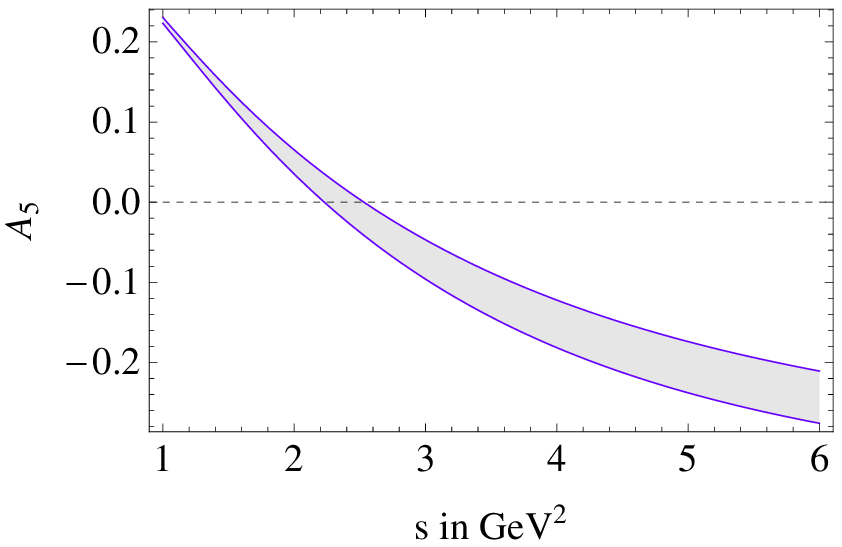}
\includegraphics[width=5cm,height=3.5cm]{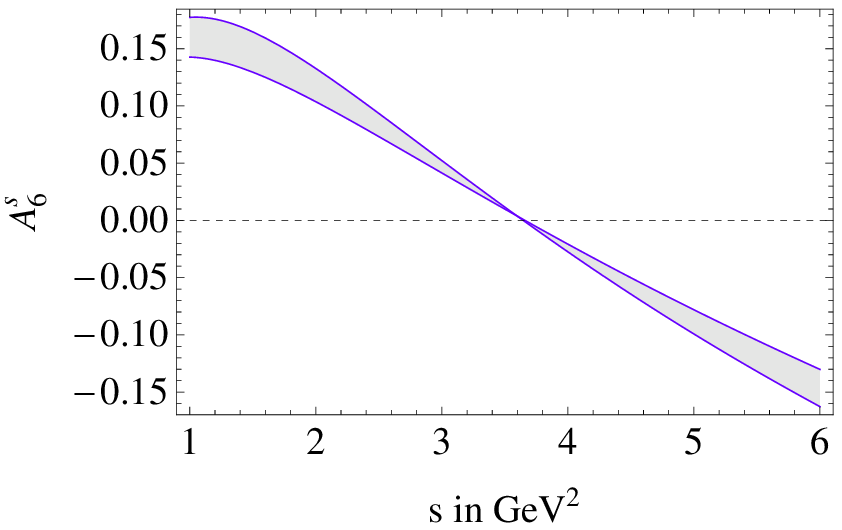}
\includegraphics[width=5cm,height=3.5cm]{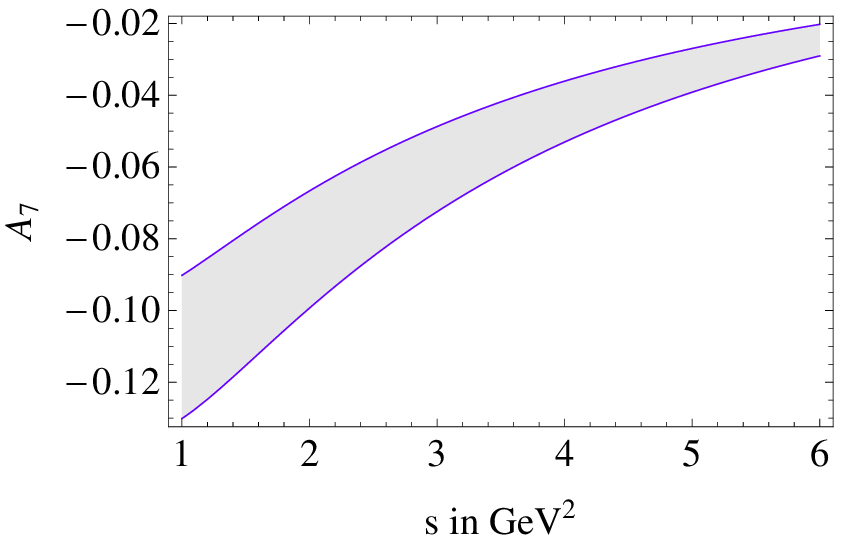}
\includegraphics[width=5cm,height=3.5cm]{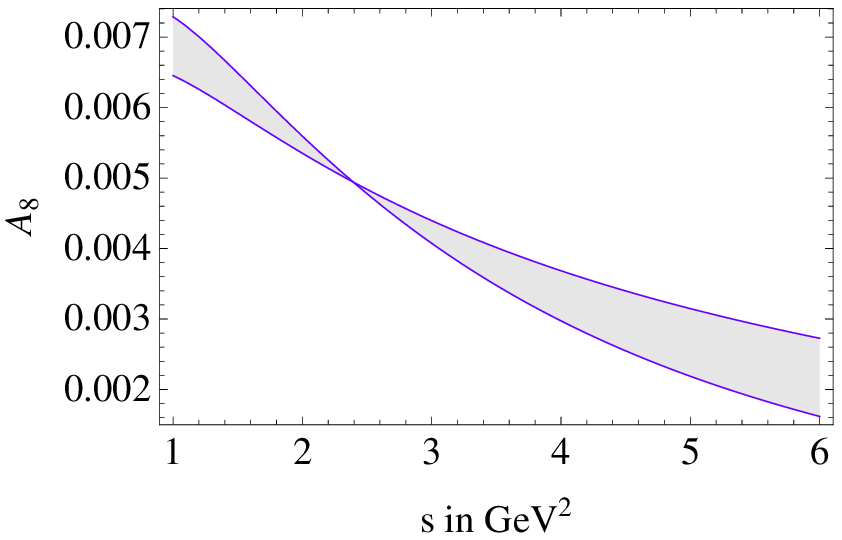}
\includegraphics[width=5cm,height=3.5cm]{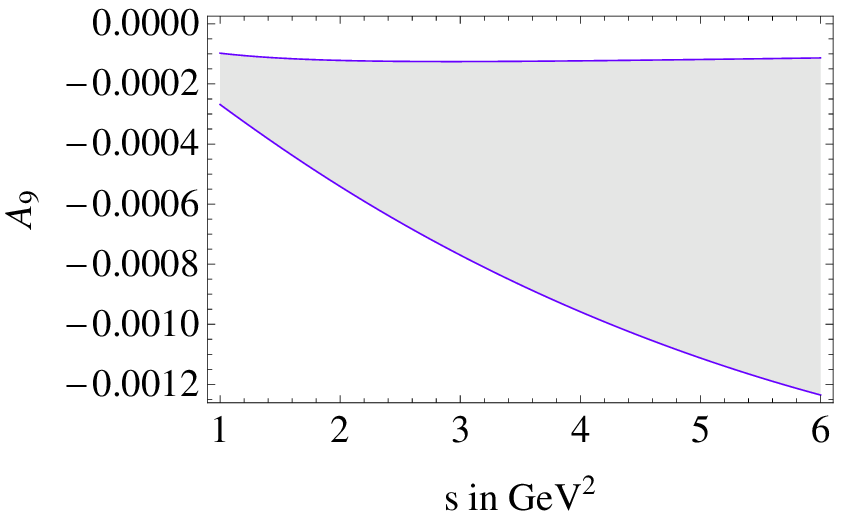}
\includegraphics[width=5cm,height=3.5cm]{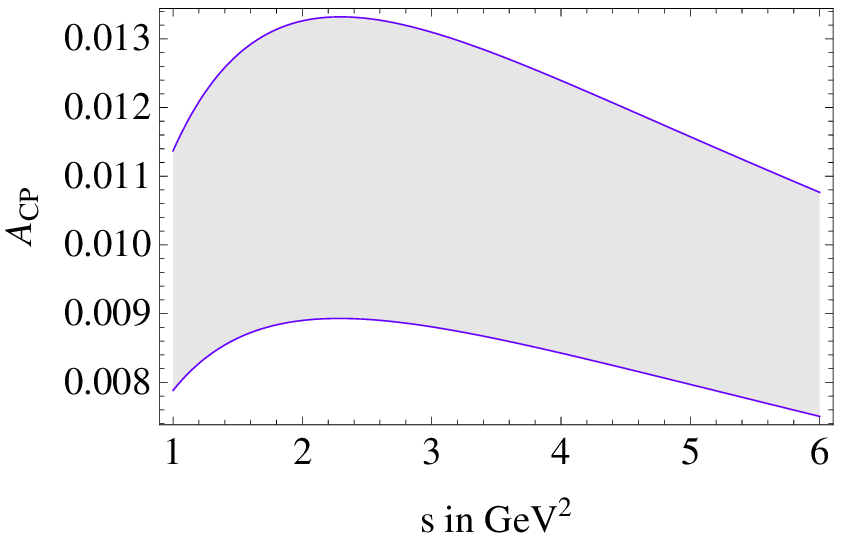}
\end{center}
\caption{Variation of the CP violating observables with di-muon invariant mass $q^2$.}
\label{cp-assym}
\end{figure}

\section{Conclusion}
In this paper we have studied the effect of the scalar leptoquarks in the rare decays of $B_s$ meson.
The large production of $B_s$ mesons at the LHC experiment opens up the possibility
to study the rare decays of $B_s$ meson with high statistical precision.
We have considered the simple renormalizable leptoquark models which do not allow
proton decay at the tree level.
Using the recent results on
${\rm BR}(B_s \to \mu^+ \mu^-)$  and the value of ${\rm BR}(\bar B_d^0 \to X_s \mu^+ \mu^-)$,
 the leptoquark parameter space has been constrained.
Using such parameters we obtained the bounds on the product of leptoquark couplings.
We then estimated the branching ratio and the
forward backward asymmetry for the rare decay process $B_s \to \phi \mu^+ \mu^-$.
The SM prediction for ${\rm BR}(B_s \to \phi \mu^+ \mu^-)$ is found to be  higher than the
corresponding experimental observed value.  We found that the branching ratio has been deviated
significantly from the corresponding SM  value and the observed branching ratio can be accommodated
in this model. However, the zero-position of the forward-backward
rate asymmetry does not have significant deviation in the leptoquark model but there is a slight shifting
 towards right. We have also shown the
variation of different CP asymmetry parameters $A_i^{(a)}$ in the low-$q^2$ region. The time-integrated
values of some of the asymmetry parameters  are found to be significantly large, the observation of which
in the LHCb experiment would provide the possible existence of leptoquarks.

{\bf Acknowledgments}

We would like to thank Council of Scientific and Industrial Research,
Government of India for financial support through grant No. 03(1190)/11/EMR-II.

\end{document}